\documentstyle[12pt]{article}
\input epsf

\def\centereps#1#2#3{\vskip0in\relax\centerline{\epsfxsize=#1\epsfysize=#2\epsfbox{#3}}}
\begin{document}
\baselineskip=24pt
\title{The Gravitational Energy of a Black Hole \\
\vspace{1.2 in} }
\author{Yuan K. Ha \\
Department of Physics, Temple University \\
Philadelphia, Pennsylvania 19122 U.S.A. \\
yuanha@temple.edu\\
\vspace{1.9 in} }
\date{General Relativity and Gravitation, Vol.35, No.11 (2003) p.2045-50}
\maketitle
\newpage
\begin{center}
\begin{large}
{\bf Abstract}
\end{large}
\end{center}
An exact energy expression for a physical black hole is derived by considering the
escape of a photon from the black hole. The mass of the black hole within its horizon
is found to be twice its mass as observed at infinity. This result is important in
understanding gravitational waves in black hole collisions.\\


%\end{abstract}

\newpage
What is the energy of a black hole? This is a question which appears
should have a simple answer. It is reasonable to conclude that the
energy of a black hole is that which corresponds to its mass as
determined by a distant observer by watching a satellite undergoing
an orbiting motion around the black hole, using the equations of
general relativity. This has been the empirical way of finding the mass
of a planet or a star. The mass obtained in this way is the total mass
of the system as seen by a distant observer. For a physical black hole, it is
the net mass obtained from the difference between the constituent mass
of the black hole and its gravitational energy. Since gravitational
energy is known to be negative, therefore the constituent mass must be
greater than the observed mass for the black hole.\\

To understand the nature of mass of a black hole, it is
necessary to know the energy distribution of the black hole throughout
all space. As the gravitational field of a black hole extends to infinity, its potential energy extends similarly and contributes also
to its observed mass. The concept of a black hole comes from the
Schwarzchild solution to Einstein's equation \cite{1}. A Schwarzchild black hole has a mass $M$ and a radius $R_{S}$ according to a distant
observer stationed at infinity.
In this paper, the total energy expression for a nonrotating
black hole including its gravitational energy is derived in a simple and physical way by considering
the escape of a photon just outside the surface of a black hole in a gedanken experiment similar to
the Hawking process \cite{2}.\\

When a photon of a given energy is emitted just outside the horizon of a black hole
it will have zero energy
as it reaches infinity. This means the entire energy of the photon is used
to escape the gravitational pull of the black hole. If the photon comes from
the annihilation of a particle of mass $m$ near the horizon, then it means the
entire mass of the particle is used to make the photon escape from the black
hole. This also means that the energy required to remove a mass $m$ just outside the
horizon to infinity is simply $mc^{2}$. Now imagine that a mass $m$ is removed
from the horizon to infinity very slowly by an external agent so that no kinetic
energy is generated in the process, the energy required to do this is still
$mc^{2}$. Eventually, the mass removed will reach infinity as a free mass.
Consider next a particle of mass $m$ being produced just outside the horizon and which has
sufficient energy to escape to infinity on its own where it ends up as a free
particle of mass $m$. The above consideration shows that the total energy required
for this event is simply $2mc^{2}$. As a result, the black hole will lose energy
by the same amount $2mc^{2}$ for each particle of mass $m$ released at the horizon and
observed at infinity. This energy is independent of the mass of the black hole.
After a succession of processes in this manner, the entire black hole is transformed
into asymptotic particles at infinity. If the total mass of the particles observed at infinity is $M$, then the original mass inside the black hole must be equal to
$2M$, half of which is used to supply the gravitational energy of these particles,
which is also the gravitational potential energy of the black hole itself.
This is a remarkable result. Thus from
the point of view of a distant observer, the constituent mass of the black hole is
$2M$, even though its observed mass is just $M$. This observed mass at infinity corresponds
to the Arnowitt-Deser-Misner mass \cite{3}, which is a measure of the total energy of a
gravitational system at spatial infinity in general relativity. A black hole thus has the maximum gravitational energy any system can have.\\

We therefore introduce the concept of the horizon mass and state the following theorem
on the energy of a black hole:
\begin{quotation}
  If {\sf M} is the mass of a black hole within its horizon,
  then its energy observed at infinity is $E$ = $\frac{1}{2}${\sf M}$c^{2}$.\\
\end{quotation}

Let us incorporate the above result into a mathematical formula. Far from the black hole, an observer should find a point mass $M$ and the
spacetime is the one described by the Schwarzchild metric.
If a photon is emitted at coordinate $r$ with energy $\varepsilon_{r}$ and
later observed at infinity, its energy there is given by
\begin{equation}
\varepsilon_{\infty} = \varepsilon_{r} \sqrt{ 1 - \frac{2GM}{rc^{2}} },
\end{equation}
where $G$ is the gravitational constant and $c$ is the speed of light.
The difference between the energies of the photon at the two locations
is therefore
\begin{equation}
\varepsilon_{r} - \varepsilon_{\infty} =
                  \varepsilon_{r}\left[ 1 - \sqrt{ 1 - \frac{2GM}{rc^{2}}} \right].
\end{equation}
The change in the photon's energy is a measure of the change of the
gravitational potential energy of the black hole as a function of
the coordinate $r$. Next, to describe the complete behavior of the energy of
the black hole itself, we introduce a function $f(r)$ interpolating between the
surface of the black hole and infinity so that the energy of the black hole also
becomes a function of the coordinate $r$. This energy expression gives the {\em total
energy} of the black hole contained in a spherical volume from the origin up to the coordinate
$r$ and is given by
\begin{equation}
E(r) = f(r)\left[ 1 - \sqrt{ 1 - \frac{2GM}{rc^{2}}} \right].
\end{equation}
To determine the function $f(r)$, we set the following conditions:\
\begin{enumerate}
\item The total energy $E(r)$ is always positive. Thus $f(r)$ must be a positive function between
      $R_{S}$ and $\infty$.
\item The total energy $E(r)$ decreases smoothly between $R_{S}$ and $\infty$. Thus its derivative
      $dE/dr$ is always negative.
\item At large distances, the total energy $E(r)$ approaches an asymptotic value.
      Thus $dE/dr \simeq 0$ at very large distances.
\end{enumerate}
Taking the derivative $dE/dr$ in Eq.(3) and subjecting it to the above conditions, we find at large
distances an equation for $f(r)$,
\begin{equation}
   \frac{df(r)}{dr} = \frac{1}{r} f(r)
\end{equation}
The solution is found to be $f(r) = {\rm constant}\times \it r.$\\

To determine the constant, we notice at large distances, the square root in Eq.(3) expands as
$\sqrt{ 1 - 2GM/rc^{2} } \simeq 1 - GM/rc^{2}$ , the energy of the black hole should approach the asymptotic value $Mc^{2}$ as seen by the distant observer. Thus,\\
\begin{equation}
  E(r) \simeq f(r) \left( {\frac{GM}{rc^{2}}} \right) \rightarrow Mc^{2},
       \hspace{.2in} r \rightarrow \infty,
\end{equation}
giving finally\\
\begin{equation}
f(r) = \frac{rc^{4}}{G}.\\
\end{equation}
The overall energy expression for the black hole is now
\begin{equation}
E(r) = \frac{rc^{4}}{G} \left[ 1 - \sqrt{ 1 - \frac{2GM}{rc^{2}} }
\right].
\end{equation}
With this result, we recover the energy of the black hole inside the Schwarzchild horizon as concluded
earlier by the distant observer. Setting $r = R_{S} = 2GM/c^{2}$, we obtain from Eq.(7)\\
\begin{equation}
E(r=R_{S}) = \left(\frac{2GM}{c^{2}}\right) \frac{c^{4}}{G} = 2Mc^{2}.
\end{equation}
The expression given by Eq.(7) agrees with the analysis of the quasilocal energy of
the Schwarzchild solution by Brown and York \cite{4}, and also agrees with the calculation
of the energy in a black hole in the teleparallel equivalent formulation of general relativity by Maluf \cite{5}. Those developments are however more mathematical and framework dependent than
the present physical approach. The significance of the present result is that the total
energy of a black hole can be found in general relativity {\em without} requiring the use of any illusive local gravitational energy density at all \cite{6}.\\

Figure 1 shows the variation of the mass of a black hole starting at $r = R_{S}$
to $r = 10 R_{S}$, using the mass equivalence of Eq.(7). As can be seen, the mass decreases quickly from $2M$ at $R_{S}$ and levels off to slightly above $M$ at $10 R_{S}$. At large distances, the mass is practically
indistinguishable from its asymptotic value $M$. However, at close distances, the mass is quite
different from $M$ as seen by the distant observer. Here the mass function is defined by
\begin{equation}
  M' = M'(r) = \frac{rc^{2}}{G} \left[ 1 - \sqrt{1 - \frac{2GM}{rc^{2}}} \right].\\
\end{equation}

\vspace{.5in}
  An important consequence of the black hole energy formula is in understanding black hole collisions.
Consider the following example.
When a black hole of asymptotic mass $5M$ collides with a black hole of asymptotic mass $12M$, the
minimum result is a black hole of asymptotic mass $13M$. This follows from the area non-decrease
theorem for black holes. The area of a black hole $A = 4\pi R^{2}_{S}$ is proportional to the
square of its asymptotic mass. Therefore, according to a distant observer watching the collision,
the amount of mass radiated away during the collision process in the form of gravitational waves
is $(5M + 12M) - 13M = 4M$.\\

  Without knowing the black hole energy formula in Eq.(7), a local observer close to the collision
process believes that the above result is always correct. This local horizon observer firmly
believes that the horizon mass is the same as the asymptotic mass because he cannot detect any
measurable changes in particle motions outside a black hole even if he were told that the horizon
mass is different from the asymptotic mass. Any particle motion is determined completely by the
Schwarzchild metric based on the asymptotic mass. Thus the local horizon observer calculates his
own orbit near the black hole based on the Schwarzchild metric and readily concludes that the
mass of the black hole is the same as when he started out from infinity. He cannot justifiably accept any other result.
But with the knowledge of the
black hole energy formula, we can understand the collision better. The collision involves a black hole
of horizon mass $10M$ with a black hole of horizon mass $24M$, resulting in a black hole of horizon
mass $26M$, again following the area non-decrease theorem. Therefore the total mass radiated away
in the collision process is $(10M + 24M) - 26M = 8M$. This is twice the amount as that concluded by
the distant observer, and also twice the amount concluded by the local horizon observer. Where has
the extra mass $4M$ gone to?\\

  If one believes that gravitational waves are responsible for the difference in mass of the black holes before
and after the collision, then this means that an additional energy of the amount $4M$ is required to allow
these waves to propagate from the final black hole to infinity for the distant observer. This is because when gravitational waves of
mass $4M$ reach infinity they will have gained potential energy of the equal amount $4M$.
Energy is inertia. The total energy
lost from the final black hole is hence $8M$, consistent with our above observation. If the local
horizon observer was correct, there would be no change in the potential energy of the gravitational waves at all.
The gravitational waves in this case cannot propagate away from the black hole.
Therefore in detecting any gravitational signal from a black hole collision such as that proposed
in the LIGO project, any conclusion about the strength of the signals near its source should be based on the
black hole energy formula. Understanding the collisions of black holes in galaxies is one of the outstanding
problems in cosmology.\\

\newpage
\vspace*{0in}
\centereps{4in}{4in}{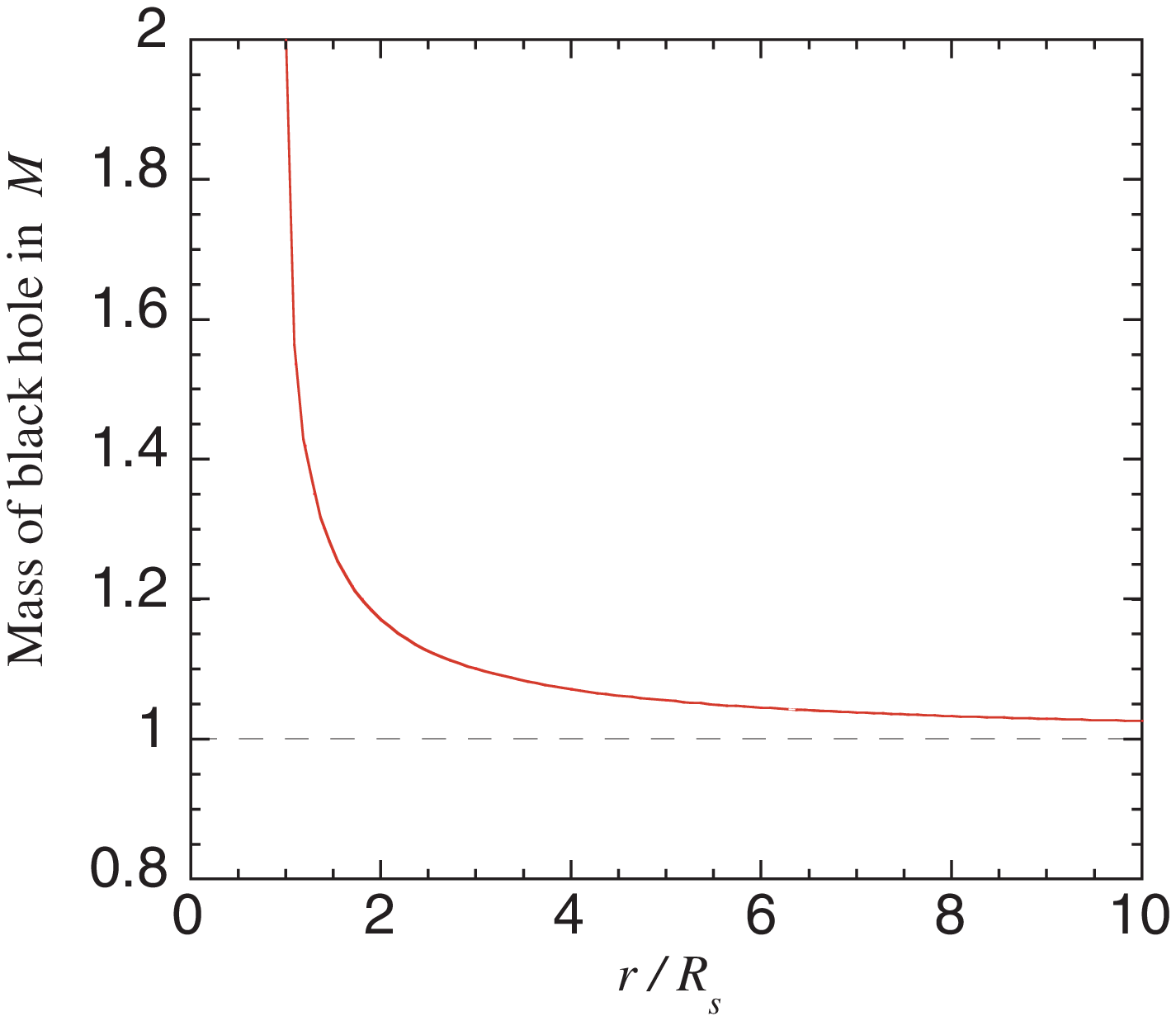}
{Figure 1. Mass of black hole as a function of radial coordinates}
\vspace{.2in}

%\begin{figure}[h]
%\begin{center}
%\epsfig{figure=blackhole.eps}
%\end{center}
%\caption{Mass of black hole as a function of radial coordinates}
%\end{figure}


\newpage
\begin{thebibliography}{99}
\bibitem{1} K. Schwarzchild, Sitzber. Deut. Akad. Wiss. Berlin, KL. Math.-Phys. Tech., (1916) 189.
\bibitem{2} S. Hawking, Commun. Math. Phys. {\bf43}, (1975) 199.
\bibitem{3} R. Arnowitt, S. Deser and C.W. Misner, in {\em Gravitation: An Introduction to Current Research}, (ed. L. Witten,  Wiley, New York 1962).
\bibitem{4} J.D. Brown and J.W. York, Jr., Phys. Rev. D{\bf47}, (1993) 1407.
\bibitem{5} J.W. Maluf, J. Math. Phys. {\bf36}, (1995) 4242.
\bibitem{6} C.C. Chang, J.M. Nester and C.M. Chen, Phys. Rev. Letts. {\bf83}, (1999) 1897.
\end{thebibliography}
\end{document}